\documentclass[letterpaper]{article}

\usepackage[T1]{fontenc}

\usepackage{geometry}
\usepackage{setspace}
\usepackage{amsmath, amssymb, graphicx, hyperref}
\usepackage{physics}  
\usepackage{graphicx}
\usepackage{enumitem}
\usepackage{booktabs}
\usepackage{epstopdf}
\usepackage{longtable}
\usepackage{epstopdf}
\DeclareGraphicsExtensions{.pdf,.png,.jpg,.eps, .tif}
\usepackage{babel}

\usepackage[super,articletitle=true]{achemso}

\usepackage{graphicx}
\usepackage{float}
\newfloat{scheme}{htbp}{los}
\floatname{scheme}{Scheme}
\floatname{chart}{Chart}
\newfloat{graph}{htbp}{loh}

\usepackage{chemformula} 
\usepackage[version = 4]{mhchem} 

\usepackage{authblk}
\author[1]{Shah Ishmam Mohtashim}
\author[2]{Manas Sajjan}
\author[1,2]{Sabre Kais*}
\affil[1]{Department of Chemistry, North Carolina State University, Raleigh, NC, USA}
\affil[2]{Department of Electrical and Computer Engineering, North Carolina State University, Raleigh, NC, USA}

\title{Continuous-time quantum-walk centrality for protein residue interaction networks}
\date{*Email: skais@ncsu.edu}

\begin{document}

\maketitle

\begin{abstract}
We present a quantum-dynamical framework for identifying structurally important residues in proteins based on continuous-time quantum walks (CTQWs) on weighted residue–interaction networks constructed from experimentally resolved structures. By mapping the weighted adjacency matrix to a Hamiltonian, residue importance emerges from the long-time averaged occupation probability, confirmed analytically through its spectral decomposition. Across a dataset of ~150 proteins spanning diverse structural and functional classes, CTQW centrality exhibits consistently strong agreement with classical eigenvector centrality in identifying central residues, while extending beyond it through incorporating signatures of quantum interference. Analyzing the time-averaged quantum transition matrix reveals consistently larger spectral gaps than the classical random-walk operator. Furthermore, biological relevance is confirmed through recovery of experimentally established functional residues in proteins kinase A and oxytocin. CTQW-derived centrality rankings are accessible on near-term intermediate-scale quantum hardware, as we demonstrate through a proof-of-principle implementation on IBM superconducting quantum hardware. These results establish continuous-time quantum walks as a computationally tractable framework for protein network analysis, that connects network theoretical treatments of protein structural biology to continuous-time quantum walk dynamics.
\end{abstract}





\section{Introduction}

Network theory provides a powerful framework for analyzing biomolecular structure and function \cite{martin1984algebraic}. In particular, residue interaction networks (RINs) recast experimentally resolved protein structures\cite{Negre2018, MOHTASHIM2025100053} as graphs, where nodes correspond to amino acid residues and edges encode spatial or energetic contacts. Over the past two decades, centrality measures on RINs have become essential tools for identifying structurally and functionally important residues, those involved in folding, stability, allosteric signaling, or ligand binding \cite{Atilgan2004, K2005-ki,DiPaola2013,Chea2007-ud,Yang2009-mh,Reetz2009-dm,Konno2019-xe,10.1093/nar/gkn433,doi:10.1021/acs.jcim.8b00146,doi:10.1073/pnas.0810961106}. Classical centrality measures, including degree, betweenness, closeness, and eigenvector centrality, have been widely used to characterize residue importance and often correlate with experimentally observed functional ``hot spots" and mutational sensitivity. For instance, residues with high betweenness centrality frequently lie along key communication pathways connecting distant regions of a protein, and their removal can significantly disrupt network connectivity\cite{TAYLOR2013e201302006}. Spectral measures such as eigenvector centrality have also been applied to identify allosteric pathways and cooperative interactions within proteins. Notably, residue centrality patterns are often more conserved across protein families than sequence-level features, highlighting their robustness as descriptors of structural organization \cite{Atilgan2004, Yang2009-mh}.

Despite their success, classical centrality measures exhibit inherent limitations. Shortest-path-based metrics, such as betweenness and closeness, emphasize geodesic routes and may overlook the presence of multiple parallel pathways that contribute to signal propagation. Conversely, spectral measures capture global influence but may fail to distinguish between structurally distinct motifs that yield similar eigenvector contributions \cite{Estrada,Bonacich1972-lm,Koschutzki2005-fn,PhysRevE.82.066102}. These challenges motivate the development of alternative frameworks that incorporate both global connectivity and the multiplicity of interaction pathways, particularly those grounded in dynamical processes on networks \cite{Estrada, PhysRevE.82.066102}. Quantum walks, the quantum-mechanical analogue of classical random walks, provide a natural extension of such dynamical approaches \cite{PhysRevA.70.022314}. Unlike classical diffusion, quantum walks evolve through coherent superposition and interference, enabling simultaneous exploration of multiple pathways across a network \cite{BerryPRL2015}. Constructive and destructive interference between these pathways can enhance or suppress contributions from specific structural features—such as hubs, cycles, or bottlenecks—offering a potentially richer characterization of node importance. 

In the continuous-time formulation, a quantum walk evolves under a Hamiltonian $H$ derived from the network structure, typically chosen as the adjacency matrix or graph Laplacian \cite{PhysRevLett.102.180501, PhysRevA.48.1687}. The resulting unitary dynamics govern the propagation of probability amplitudes across the network, with transition probabilities reflecting underlying connectivity. Centrality measures derived from time-averaged occupation probabilities of continuous-time quantum walks (CTQWs) have been shown to identify important nodes in complex and weighted graphs, often with enhanced sensitivity compared to classical approaches \cite{Paparo2012,Wang2022,Wang:20,PhysRevLett.125.240501}. Recent advances in quantum computing further enable the direct realization of quantum-walk dynamics on programmable quantum devices \cite{kendon2020quantum}. In this setting, the network adjacency matrix serves as a Hamiltonian whose evolution operator $e^{-iHt}$ can be implemented through quantum circuits, with this capability providing a promising route toward performing network analysis using quantum resources, particularly as quantum hardware continues to scale \cite{decision, PhysRevA.70.022314}. To date, however, the application of quantum-walk-based centrality to protein residue interaction networks has remained unexplored. 

Previous work has approached residue centrality from a quantum perspective using analogue quantum optimization-based formulations, for instance mapping to a quadratic unconstrained binary optimization (QUBO) model \cite{MOHTASHIM2025100053} solvable on quantum annealers \cite{camino2023quantum}. In contrast, we introduce a dynamical framework in which the weighted adjacency matrix of a RIN generates quantum-walk dynamics, and residue importance is defined through the long-time averaged occupation probability of a given quantum walk. From a computational standpoint, CTQW centrality corresponds to analyzing the Schr\"odinger evolution generated by a typically sparse Hamiltonian defined by the interaction network. As continuous-time quantum walks intrinsically incorporate global connectivity and contributions from multiple pathways through interference, they provide a dynamical generalization of classical spectral centrality \cite{ambainis03, Loke2016, PhysRevA.48.1687}. In the context of proteins, where structural communication often involves redundant pathways and cooperative interactions \cite{liu2001quantum, okazaki2008dynamic, protein_motion}, this perspective offers a principled means of resolving features that may not be captured by classical metrics alone.

We develop a unified quantum-walk-based framework for protein network analysis, supported by an exact analytic expression for the steady-state distribution of CTQWs via spectral decomposition, thus enabling efficient and parameter-free quantum computation of centrality. We benchmark the resulting centrality measures against classical eigenvector centrality across a diverse dataset of proteins and demonstrate strong agreement in identifying structurally important residues. Finally, we present a proof-of-principle quantum-circuit implementation on IBM superconducting hardware, illustrating the feasibility of evaluating quantum-walk-based centrality on near-term devices.

\section{Residue Interaction Networks}

Residue Interaction Networks (RINs) were constructed from Protein Data Bank (PDB)\cite{10.1093/nar/28.1.235} structures using the Biopython Bio.PDB module.\cite{Biopython} Each amino acid residue C-$\alpha$ atom was represented as a node, and edges were defined based on spatial proximity between residues. Specifically, two residues $i$ and $j$ were considered connected if the Euclidean distance between their C-$\alpha$ atoms, denoted $d_{ij}$, satisfied the cutoff criterion $d_{ij} < 8~\text{\AA}$. This $8~\text{\AA}$ threshold follows standard conventions in protein network analysis and provides a coarse-grained yet informative mapping of tertiary structure.\cite{Negre2018, K2005-ki}

To capture the geometric interaction strength between neighboring residues, each edge $(i,j)$ was assigned a weight inversely proportional to the squared C-$\alpha$ distance:
\begin{equation}
w_{ij} = \frac{1}{(d_{ij})^2},
\end{equation}
where $d_{ij}$ is expressed in \AA ngstr\"{o}ms, so that $w_{ij}$ carries units of \AA$^{-2}$. The adjacency matrix $A$ therefore inherits these units, and since $H = A$ is used as the Hamiltonian with $\hbar = 1$, the evolution time $t$ in the Schr\"{o}dinger equation carries units of \AA$^{2}$. Closer residues contribute more strongly to the interaction network, and the resulting graph is an undirected, weighted network $G = (V, E)$, where $V$ denotes the set of residues and $E$ represents pairwise spatial interactions below the cutoff threshold.

The adjacency matrix $A$ of each network was subsequently used as the Hamiltonian $H$ in the continuous-time quantum walk simulations. All node numbering in Figure 1 directly follows the author residue numbering provided in the original PDB file, ensuring consistency with experimentally determined structures. This preserves the canonical residue indices used by the depositor, allowing direct correspondence between graph nodes and protein residues. 

 \begin{figure}[t]
  \centering
  \includegraphics[width=0.7\linewidth]{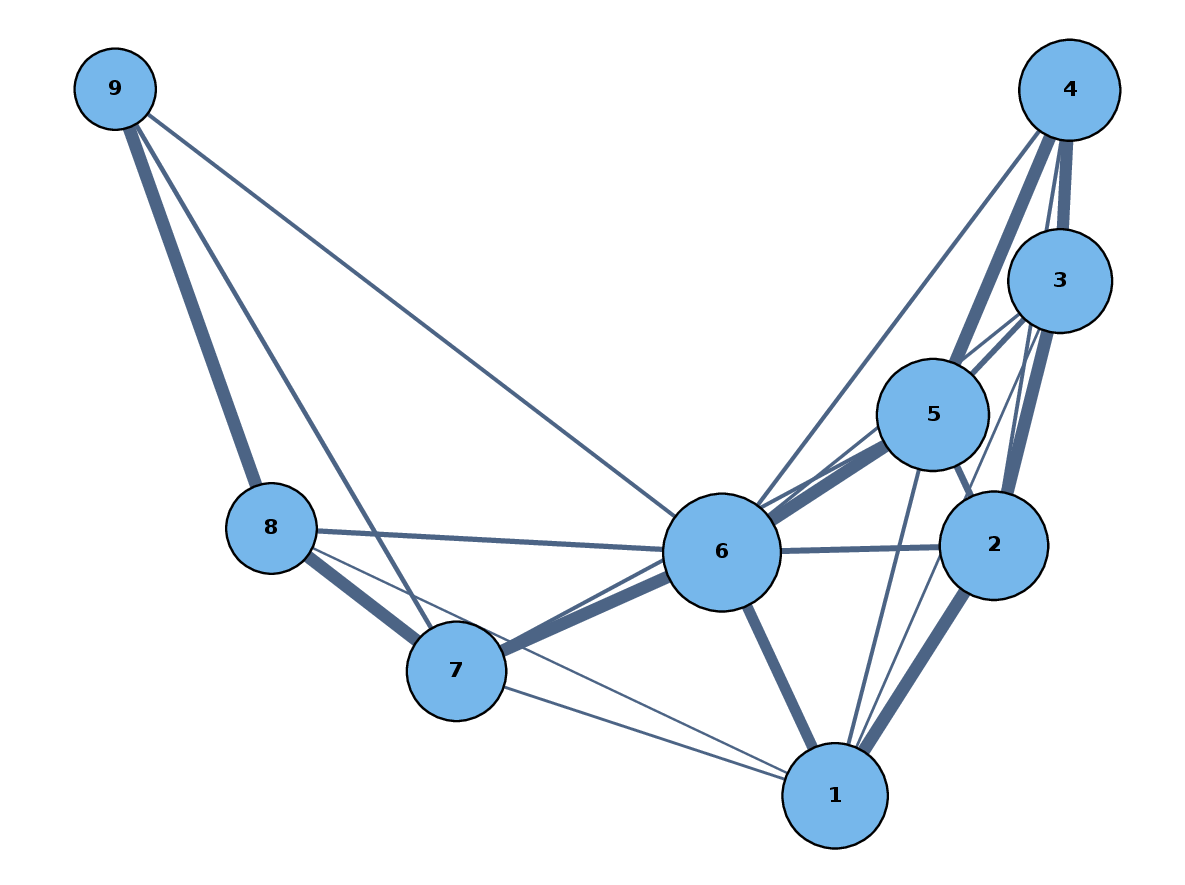}
  \caption{Residue--interaction network for oxytocin (PDB ID: 1XY1). Nodes represent residues, with node size proportional to their eigenvector centrality. Edges denote C\textsubscript{$\alpha$} contacts under an 8\,\AA\ cutoff, with edge thickness scaled according to the inverse-square distance weighting between residues.}
  \label{fig:rin-oxytocin}
\end{figure}

\section{Classical Baseline - Eigenvector Centrality}
Given a weighted residue–interaction network with adjacency matrix $A\in\mathbb{R}^{n\times n}$, the classical eigenvector centrality assigns to each residue $i$ a score $x_i$ proportional to the centrality of its neighbors. In vector form,
\begin{equation}
A\,\mathbf{x}=\lambda_{\max}\,\mathbf{x}, \qquad \mathbf{x}\ge 0,\quad \
\label{eq:eigencent}
\end{equation}
where $\lambda_{\max}$ is the largest eigenvalue of $A$ and the nonnegativity follows from the Perron–Frobenius theorem.\cite{Bonacich1972-lm, Estrada, https://doi.org/10.1002/nav.22042} For our RINs, $A$ is built from C$_\alpha$ contact weights (e.g., $w_{ij}=1/(d_{ij})^2$ under an 8\,\AA\ cutoff), so \eqref{eq:eigencent} measures residues that are strongly coupled to other influential residues. In practice we compute $\mathbf{x}$ with the power method as implemented in NetworkX eigenvector\_centrality \cite{NetworkX,Hagberg2008-wq}, using a uniform initialization, tolerance $10^{-6}$, and at most $10^3$ iterations.


\section{Continuous-Time Quantum-Walk Centrality}

\subsection{Continuous-Time Quantum Walk}
A continuous-time quantum walk (CTQW) evolves on a graph $G$ under a time-independent Hamiltonian that reflects the underlying network topology.  For an undirected graph with $n$ vertices, the walker occupies a Hilbert space spanned by the orthonormal basis $\{|1\rangle, |2\rangle, \ldots, |n\rangle\}$, each state corresponding to a vertex of the graph. The quantum dynamics follow the Schr\"odinger equation
\begin{equation}
    i\hbar \frac{d}{dt} |\psi(t)\rangle = H |\psi(t)\rangle,
    \label{eq:schrodinger}
\end{equation}
where the Hamiltonian $H$ is identified with the adjacency matrix $A$ of the network. For weighted graphs \cite{chatterjee_2026}, the matrix elements $A_{ij}$ denote the edge  weights between nodes $i$ and $j$:
\begin{equation}
A_{ij} = 
\begin{cases}
w_{ij}, & \text{if nodes $i$ and $j$ are connected,}\\[3pt]
0, & \text{otherwise.}
\end{cases}
\label{eq:adjacency}
\end{equation}
Solving Eq.~(\ref{eq:schrodinger}) with $\hbar = 1$ yields the time-evolved 
state of the walker,
\begin{equation}
    |\psi(t)\rangle = e^{-iHt} |\psi(0)\rangle 
    = \sum_{i=1}^{n} \alpha_i(t)\, |i\rangle,
    \label{eq:ctqw_state}
\end{equation}
where $\alpha_i(t) = \langle i|\psi(t)\rangle$ is the probability amplitude at 
vertex $i$ and $P_i(t) = |\alpha_i(t)|^2$ the corresponding occupation probability.  

For centrality analysis, the walk begins from a uniform superposition over all vertices, $|\psi(0)\rangle = \frac{1}{\sqrt{n}}\sum_{i=1}^{n} |i\rangle$, allowing the walker to explore the graph in coherent superposition.  
The CTQW-based centrality of node $i$ is then defined as the long-time average of its occupation probability,  
\begin{equation}
    c_i^{(\mathrm{CTQW})} 
    = \bar P_i 
    = \lim_{T \to \infty} \frac{1}{T} 
      \int_0^{T} |\langle i|\psi(t)\rangle|^2\, dt.
    \label{eq:ctqw_centrality}
\end{equation}
This formalism establishes a direct connection between network connectivity and 
quantum propagation dynamics, providing a coherent, interference-driven analogue 
of classical spectral centrality measures~\cite{PhysRevA.70.022314,Paparo2012,Wang2022}.

\subsubsection{Choice of Hamiltonian: Adjacency vs.\ Laplacian}

In the CTQW formalism, the Hamiltonian $H$ is commonly chosen as either the adjacency matrix $A$ or the graph Laplacian $L = D - A$, where $D_{ii} = \sum_j A_{ij}$.\cite{PhysRevA.70.022314, PhysRevLett.116.100501} Because we seek a global, unbiased ranking of all residues simultaneously, the walker is initialized in a uniform superposition,
\begin{equation}
\ket{\psi(0)} = \ket{s} = \frac{1}{\sqrt{n}} \sum_{i=1}^n \ket{i},
\end{equation}
allowing the dynamics to differentiate residues purely through network connectivity.

For this initialization, the Laplacian yields trivial dynamics: $\ket{s}$ is a zero-eigenvalue eigenstate of $L$, since $L\ket{s} = (D-A)\ket{s} = 0$, and consequently $e^{-iLt}\ket{s} = \ket{s}$ for all $t$. The occupation probabilities remain uniform and encode no structural information. Although localized initial states would restore nontrivial Laplacian dynamics, they introduce bias toward the starting node and require a separate computation per residue or an additional averaging step over all starting sites, substantially increasing cost without clear physical motivation.

In contrast, the adjacency matrix does not admit $\ket{s}$ as an eigenvector for irregular or weighted graphs such as protein RINs. Evolution under $H = A$ therefore produces nontrivial coherent dynamics driven by interference between multiple interaction pathways, and the resulting time-averaged occupation probabilities reflect the global connectivity of the network in a single computation. For these reasons, all CTQW simulations in this work employ the weighted adjacency matrix as the Hamiltonian.

\subsubsection{Analytic Steady-State of the CTQW formulation}
\label{sec:ctqw_steady_state}
The dynamics of a continuous-time quantum walk (CTQW) on a graph with Hamiltonian $H$ are governed by
\begin{equation}
    |\psi(t)\rangle = U(t)|\psi_0\rangle,
    \qquad
    U(t) = e^{-iHt}.
\end{equation}

The probability of observing the walker at node $i$ is
\begin{equation}
    P(i,t \mid \psi_0)
    =
    |\langle i | \psi(t) \rangle|^2
    =
    \langle i | U(t) | \psi_0 \rangle
    \langle \psi_0 | U^\dagger(t) | i \rangle .
\end{equation}

Let $\{|\lambda_m\rangle\}$ denote the eigenbasis of $H$ with eigenvalues $\{\lambda_m\}$,
\begin{equation}
    H = \sum_m \lambda_m |\lambda_m\rangle\langle \lambda_m|.
\end{equation}

Expanding the initial state in this basis,
\begin{equation}
    |\psi_0\rangle = \sum_m c_m |\lambda_m\rangle,
    \qquad
    c_m = \langle \lambda_m | \psi_0 \rangle,
\end{equation}
the transition amplitude becomes
\begin{equation}
    \langle i | \psi(t) \rangle
    =
    \sum_m c_m \langle i | \lambda_m \rangle e^{-i\lambda_m t}.
\end{equation}

The probability can therefore be written as a double spectral sum,
\begin{equation}
    P(i,t \mid \psi_0)
    =
    \sum_{m,n}
    c_m c_n^*
    \langle i | \lambda_m \rangle
    \langle \lambda_n | i \rangle
    e^{-it(\lambda_m - \lambda_n)}.
\end{equation}

To define a steady-state distribution, we consider the long-time average
\begin{equation}
    \bar{P}(i)
    =
    \lim_{T \to \infty}
    \frac{1}{T}
    \int_0^T
    P(i,t \mid \psi_0)\, dt .
\end{equation}

Substituting the spectral expansion,
\begin{equation}
    \bar{P}(i)
    =
    \sum_{m,n}
    c_m c_n^*
    \langle i | \lambda_m \rangle
    \langle \lambda_n | i \rangle
    \lim_{T \to \infty}
    \frac{1}{T}
    \int_0^T
    e^{-it(\lambda_m - \lambda_n)} dt .
\end{equation}

The time integral evaluates to
\begin{equation}
    \lim_{T\to\infty}
    \frac{1}{T}
    \int_0^T e^{-it(\lambda_m-\lambda_n)} dt
    =
    \begin{cases}
        1, & \lambda_m = \lambda_n, \\
        0, & \lambda_m \neq \lambda_n.
    \end{cases}
\end{equation}

Thus, all oscillatory terms with $\lambda_m \neq \lambda_n$ vanish in the long-time limit, and only terms satisfying $\lambda_m=\lambda_n$ survive.The steady-state distribution becomes
\begin{equation}
    \bar{P}(i)
    =
    \sum_{m,n:\,\lambda_m=\lambda_n}
    c_m c_n^*
    \langle i | \lambda_m \rangle
    \langle \lambda_n | i \rangle.
\end{equation}


In the absence of spectral degeneracies, this condition is satisfied only when $m=n$, since all eigenvalues are distinct. The steady-state distribution therefore reduces to
\begin{equation}
    \bar{P}(i)
    =
    \sum_m
    |c_m|^2
    |\langle i | \lambda_m \rangle|^2.
\end{equation}

This result follows because terms with $\lambda_m \neq \lambda_n$ carry oscillatory phases $e^{-i(\lambda_m-\lambda_n)t}$ that average to zero over long times, eliminating all interference between different eigenmodes. When degeneracies are present, however, distinct eigenstates sharing the same eigenvalue do not accumulate relative phase, and additional off-diagonal contributions within those eigenspaces persist.

The quantity $\bar{P}(i)$ thus represents the infinite-time averaged probability of finding the walker at node $i$, providing a closed-form description of the asymptotic behavior of the quantum walk without requiring explicit long-time simulation. Accordingly, we define the CTQW centrality score of node $i$ as
\begin{equation}
    c_i^{(\mathrm{CTQW})} = \bar{P}(i).
\end{equation}

\subsubsection{Scalability and Spectral Estimation via FFT}

While the infinite-time steady state $\bar{P}(\infty)$ can be obtained analytically from the spectral decomposition of the adjacency matrix $A$, this approach becomes computationally demanding for large networks. For a residue interaction network with $N$ residues, the Hamiltonian is
\begin{equation}
A \in \mathbb{R}^{N \times N},
\end{equation}
and computing $\bar{P}(\infty)$ requires diagonalizing $A$. For dense matrices, exact eigendecomposition scales as $\mathcal{O}(N^3)$ in time, which can become prohibitive for large proteins or ensemble studies involving hundreds of structures.

In large residue interaction networks, explicit diagonalization of the adjacency matrix $A$ is computationally prohibitive, making the analytic spectral pinching limit inaccessible. Consequently, one cannot directly verify whether the long time average mean $\bar{P}(T)$ has converged to $\bar{P}(\infty)$.

To address this, one may use a frequency-domain diagnostic based on the Fast Fourier Transform. For each residue $i$, the sampled probability signal $P_i(t_k)$ is treated as a discrete time series and transformed via
\begin{equation}
X_i(f_m)
=
\sum_{k=0}^{L-1}
P_i(t_k)\, e^{-2\pi i m k/L}.
\end{equation}
Under the FFT convention, the zero-frequency bin satisfies
\begin{equation}
X_i(0)
=
\sum_{k=0}^{L-1} P_i(t_k),
\end{equation}
so that the DC component is
\begin{equation}
\mathrm{DC}_i
=
\frac{\operatorname{Re} X_i(0)}{L}.
\end{equation}
Since the infinite-time average
\begin{equation}
\bar{P}_i
=
\lim_{T\to\infty}
\frac{1}{T}\int_0^T P_i(t)\,dt
\end{equation}
corresponds precisely to the zero-frequency contribution of the signal, stabilization of the DC component provides a practical heuristic of steady-state behavior.

Operationally, one can compute the DC vector over sliding time windows and monitor the window-to-window variation
\begin{equation}
\Delta_k
=
\|\mathbf{d}^{(k+1)} - \mathbf{d}^{(k)}\|_\infty.
\end{equation}

The steady time $t^{\*}$ is defined as the earliest time at which $\Delta_k$ falls below a prescribed tolerance for several consecutive windows. This FFT-based approach thus enables steady-state detection without explicit spectral decomposition of $A$.

\subsubsection{Spectral Gap Analysis}

To characterize the structural signatures encoded by classical diffusion and coherent quantum transport, we analyze the spectral properties of their respective transition operators.

\paragraph{Classical random walk.}
Given the weighted adjacency matrix $A$, we construct the row-stochastic transition matrix
\begin{equation}
P = D^{-1}A,
\end{equation}
where $D = \mathrm{diag}(d_1,\dots,d_n)$ and $d_i = \sum_j A_{ij}$ is the weighted degree of node $i$. The stationary distribution $\pi$ satisfies $\pi^T P = \pi^T$, and for connected graphs is given by
\begin{equation}
\pi_i = \frac{d_i}{\sum_k d_k}.
\end{equation}
The classical spectral gap, which governs the rate of convergence to the stationary distribution under repeated application of $P$, is defined as
\begin{equation}
\Delta_{\mathrm{cl}} = 1 - \max_{k \ge 2} |\lambda_k(P)|.
\end{equation}

\paragraph{Quantum walk: time-averaged transition matrix.}
Because the unitary evolution governing a CTQW preserves state purity and does not converge to a fixed distribution, we instead construct an effective transition matrix from the infinite-time averaged transition probabilities between all pairs of nodes:
\begin{equation}
T^{(Q)}_{ij}
=
\lim_{T \to \infty}
\frac{1}{T}
\int_0^T
\left|
\langle i | e^{-iAt} | j \rangle
\right|^2
dt.
\end{equation}
Using the spectral decomposition $A = \sum_\alpha \lambda_\alpha \Pi_\alpha$, where $\Pi_\alpha$ projects onto the eigenspace associated with eigenvalue $\lambda_\alpha$, this limit can be evaluated analytically as
\begin{equation}
T^{(Q)}_{ij} = \sum_\alpha \left| \langle i | \Pi_\alpha | j \rangle \right|^2.
\end{equation}
In particular, $T^{(Q)}$ coincides with the infinite-time mean of the quantum transition probabilities, reflecting dephasing in the energy eigenbasis. The resulting matrix is nonnegative and row-stochastic. We define the quantum spectral gap analogously:
\begin{equation}
\Delta_Q = 1 - \max_{k \ge 2} |\lambda_k(T^{(Q)})|.
\end{equation}

The spectral gap provides a natural timescale for convergence: the mixing time is inversely proportional to the gap, $t_{\mathrm{mix}} \sim \Delta^{-1}$.\cite{PhysRevA.102.022423} Accordingly, we use $\Delta_{\mathrm{cl}}^{-1}$ and $\Delta_Q^{-1}$ as effective measures of mixing times to enable a direct comparison between classical diffusion and the quantum walk. To ensure that both gaps reflect network structure rather than graph reducibility, all computations are performed on the largest connected component of each RIN.

\section{Quantum circuit implementation}
\label{subsec:ctqw_pl}

\paragraph{Objective.}
Given a weighted, undirected residue graph with adjacency \(A\in\mathbb{R}^{n\times n}\), the continuous-time quantum walk (CTQW) centrality for residue \(i\) is defined as the long-time average of the visit probability under the unitary dynamics generated by \(A\):
\begingroup
\setlength{\jot}{2pt}\setlength{\arraycolsep}{2pt}
\begin{equation}
\label{eq:ctqw_cent_def}
\begin{aligned}
\bar P_i &= \lim_{T\to\infty}\frac{1}{T}\int_{0}^{T}\!\abs{\braket{i}{\psi(t)}}^{2}\,dt,\\
\ket{\psi(t)}&= e^{-iAt}\ket{\psi_0},\qquad
\ket{\psi_0}=\tfrac{1}{\sqrt{n}}\sum_{i=1}^{n}\ket{i}.
\end{aligned}
\end{equation}
\endgroup
In practice, this long-time average is approximated by a finite Cesàro mean over a discrete grid of evolution times \(\mathcal{T}=\{t_\ell\}\).\cite{Paparo2012,Wang:20,PhysRevLett.125.240501}

\paragraph{Register padding and initial state.}
To map the \(n\)-dimensional residue space onto a qubit register, the Hilbert space is padded to the nearest power of two. Let \(q=\lceil\log_2 n\rceil\) and \(N=2^q\), where $n$ is the number of residues and $q$ is the number of qubits. The adjacency matrix is embedded into an \(N\times N\) Hamiltonian
\[
A_p=\begin{bmatrix}A&0\\[2pt]0&0\end{bmatrix},
\]
where the lower-right zero block ensures that the additional basis states remain dynamically inert. The initial quantum state is prepared as a uniform superposition over the physical residues only,
\[
\ket{\psi_0}=\frac{1}{\sqrt{n}}\sum_{i=0}^{n-1}\ket{i},
\]
assigning zero amplitude to the padded states. The dynamics therefore remain confined to the physical subspace, since
\[
e^{-iA_p t}=\mathrm{diag}(e^{-iAt},\,I_{N-n}).
\]
\paragraph{Qubit Scaling with Residue Number}

The number of qubits required to represent a protein network grows only  logarithmically with the number of residues. A quantum register of $q$ qubits  can encode $2^{q}$ distinct basis states, each corresponding to a residue node in the network. Therefore, the minimum qubit requirement is 
$q = \lceil \log_{2} n \rceil$, where $n$ is the number of residues. For example, the oxytocin network ($n=9$) requires $q=4$ qubits, providing  $2^{4}=16$ basis states. The first nine basis states correspond to physical  residues, while the remaining seven are inert padding states that remain unpopulated during evolution. Oxytocin, with nine residues, requires only four qubits, while mid-sized proteins such as ubiquitin ($residues=76$), thioredoxin ($residues=108$), and PDZ domains ($residues=94$) fit within seven qubits. Larger systems like lysozyme ($residues=129$) or a 300-residue enzyme can be represented on eight to nine qubits, and even a 512-residue protein would still require only nine qubits. This compact logarithmic scaling highlights the exceptional resource efficiency of the CTQW encoding, enabling simulations of biologically relevant proteins on near-term quantum devices.

\paragraph{Time evolution as a circuit.}
For each evolution time \(t\in\mathcal{T}\), the state is evolved by the exact unitary operator \(U(t)=e^{-iA_p t}\) on \(q\) qubits, and the probability distribution is read out in the computational basis:
\begingroup
\setlength{\jot}{2pt}\setlength{\arraycolsep}{2pt}
\begin{equation}
\label{eq:ctqw-circuit}
\begin{aligned}
\text{(prep)}\;& \ket{0}^{\otimes q}\xrightarrow{\;\mathrm{M\ddot{o}tt\ddot{o}nen}\;}\ket{\psi_0},\\
\text{(evol)}\;& \ket{\psi(t)} = U(t)\ket{\psi_0},\quad U(t)=e^{-iA_p t},\\
\text{(readout)}\;& p_i(t)=|\langle i|\psi(t)\rangle|^2,\;\; \mathbf{p}(t)=(p_0,\ldots,p_{N-1}).
\end{aligned}
\end{equation}
\endgroup

In PennyLane, \(\ket{\psi_0}\) is prepared using MottonenStatePreparation,\cite{motten} and \(U(t)\) is applied via a single QubitUnitary operation constructed from the dense matrix exponential \(\mathrm{expm}(-iA_p t)\). \cite{PennyLane2022}

 \begin{figure}[t]
  \centering
  \includegraphics[width=0.7\columnwidth,trim=10 20 10 15,clip]{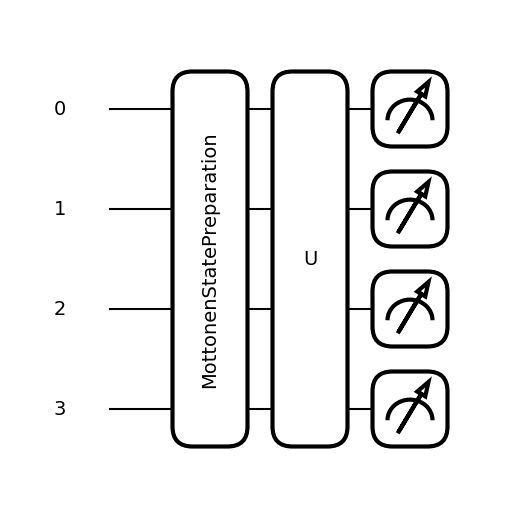}
\caption{Quantum circuit for CTQW centrality of oxytocin (PDB ID: 1XY1).
Four qubits encode the $n=9$ residues plus seven inert padding states.
\textsc{MottonenStatePreparation} initializes the uniform superposition
$|\psi_0\rangle = \frac{1}{\sqrt{9}}\sum_{i=0}^{8}|i\rangle$, and $U =
e^{-iA_p t}$ applies the padded adjacency Hamiltonian for evolution time
$t$. Computational-basis measurement yields $p_i(t) = |\langle
i|\psi(t)\rangle|^2$; averaging over a time grid $\mathcal{T}$ produces
the centrality scores.}
  \label{fig:rin-circuit}
\end{figure}

\paragraph{Time averaging and truncation.}
The probabilities corresponding to the physical residues are extracted by retaining the first \(n\) components of \(\mathbf{p}(t)\) and discarding the padded entries. The CTQW centrality vector is then obtained by averaging these over all time points:
\[
c^{(\mathrm{CTQW})}_i
=\frac{1}{|\mathcal{T}|}\sum_{t\in\mathcal{T}} p_i(t),
\qquad i=0,\dots,n-1.
\]
Unless otherwise specified, we use a uniform grid of \(L=40\) time points up to \(20\pi\),
\[
\mathcal{T}=\{t_\ell\}_{\ell=1}^{L}, \quad t_\ell\in[0,\,20\pi],
\]
which follows standard practice in CTQW centrality analyses, yielding a numerically stable estimator of Eq.~\eqref{eq:ctqw_cent_def}. The resulting vector is renormalized to ensure unit total probability after truncation.

\paragraph{Computational Complexity and Practical Considerations}
For a protein with $n$ residues, directly forming the dense matrix exponential
$e^{-iAt}$ by eigendecomposition or generic dense linear algebra is typically
$\mathcal{O}(n^{3})$ in time and $\mathcal{O}(n^{2})$ in memory.
This becomes a bottleneck for large graphs or repeated evaluations over many
time points.

On a quantum computer, the unitary evolution $e^{-iHt}$ for a \emph{sparse}
Hamiltonian $H$ can be implemented using sparse-Hamiltonian simulation
algorithms with complexity that is polylogarithmic in the Hilbert-space
dimension under standard oracle access models, with an explicit dependence on
the sparsity $s$, evolution time $t$, and target error $\epsilon$
(e.g., scaling of the form $\mathrm{poly}\!\big(\log n,\, s,\, t,\, \log(1/\epsilon)\big)$,
up to method-dependent factors)~\cite{BerryPRL2015, 10.21468/SciPostPhysCore.6.3.058}.
This suggests an asymptotically favorable route for simulating CTQW
dynamics on large residue-interaction networks, which are typically sparse.

We emphasize, however, that realizing such asymptotic advantages requires
(i) efficient access to the sparse matrix elements, and (ii) accounting for the costs of state preparation and measurement/readout, which can dominate in near-term settings. In the present work, our hardware demonstration uses a proof-of-principle dense-unitary compilation of $U(t)$ QubitUnitary, which is convenient for small $n$ but does not exploit sparsity and does not reflect the scaling of advanced Hamiltonian-simulation
methods. Future implementations should employ structure-preserving sparse simulation techniques (e.g., qubitization-based methods) to reduce circuit depth while maintaining the same CTQW averaging protocol.

\begin{figure*}[t]
  \centering
  \includegraphics[width=0.90\textwidth]{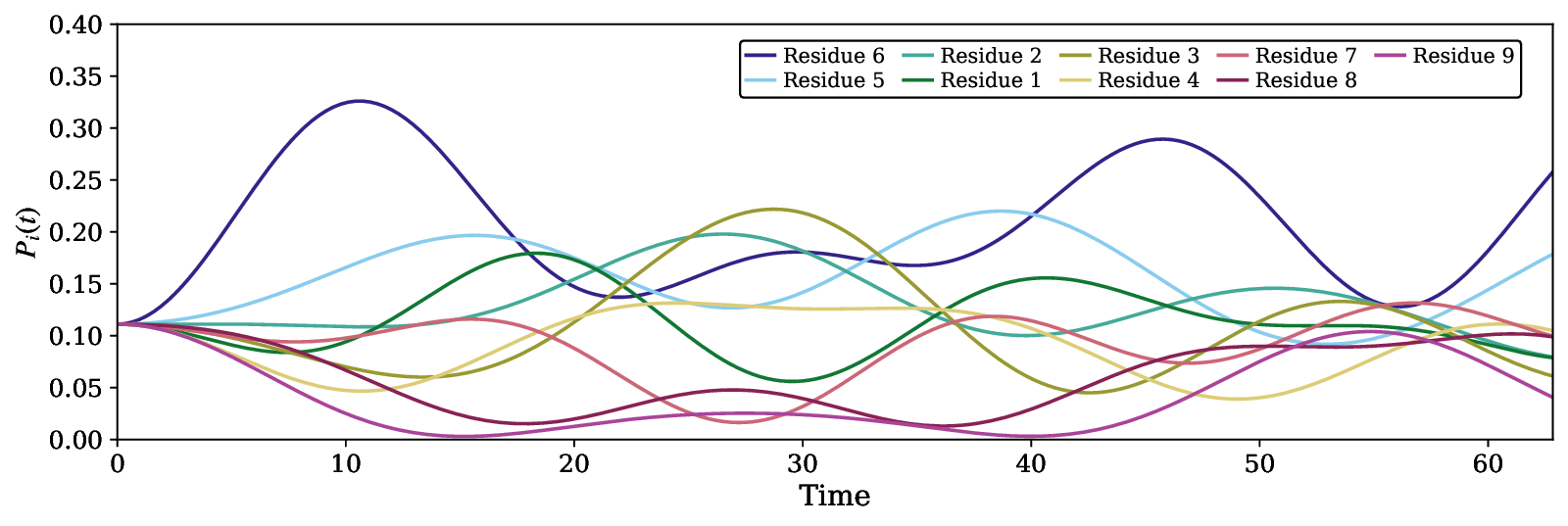}
  \caption{Time evolution of node-occupation probabilities 
  $P_i(t)=\big|\!\braket{i}{\psi(t)}\!\big|^2$ for residues ranked by CTQW centrality.}
  \label{fig:probtime}
\end{figure*}

\paragraph{Probability Distribution with Time}
Beyond time-averaged centrality, we can also track the full probability distribution $P_i(t)$ for each residue across the evolution time $t$. This dynamic profile provides additional insights into transient localization and spreading phenomena, offering a richer characterization of residue importance than static measures. This is shown in Figure \ref{fig:probtime}.

\section{Dataset and evaluation metrics}

\subsection{Protein Dataset}

We analyzed a broad panel of around 150 proteins spanning ten structural and functional classes to evaluate the generality of CTQW-based centrality across diverse residue interaction topologies. The dataset includes the following classes of proteins: 

\begin{enumerate}[label=(\Alph*)]
    \item Classic folding and stability benchmarks 
    \item DNA-binding and transcription-factor domains 
    \item RNA-binding domains 
    \item Signalling modules 
    \item Immunoglobulin-like and adhesion modules
    \item Metalloproteins
    \item Small catalytic enzymes
    \item Toxins and antimicrobial peptides
    \item Membrane proteins and helical bundles
    \item Hormones and growth factors
\end{enumerate}

The comprehensive list of proteins, their corresponding PDB identifiers and their residue numbers is provided in Supplementary Information 1.

\subsection{Correlation Metrics }
To quantify the agreement between the eigenvector and CTQW centrality rankings, we employ four complementary metrics: Spearman’s rank correlation~$\rho$, Kendall’s tau~$\tau$, Overlap, and the Jaccard index.\cite{Spearman, 10.1093/biomet/30.1-2.81} These jointly capture rank-order and set-level similarities between the two methods.

\subsubsection{Rank-Based Correlation Measures}

\paragraph*{Spearman’s rank correlation ($\rho$).}
Spearman’s $\rho$ is the Pearson correlation of rank variables with tie correction:
\begin{align}
\rho &=
\frac{
\sum_i (x_i - \bar{x})(y_i - \bar{y})
}{
\sqrt{\sum_i (x_i - \bar{x})^2}\;
\sqrt{\sum_i (y_i - \bar{y})^2}
}.
\end{align}

\paragraph*{Kendall’s tau ($\tau$).}
To quantify pairwise ordering consistency between two rankings, we compute Kendall’s rank correlation coefficient $\tau$, which accounts for tied values. Let $C$ and $D$ denote the numbers of concordant and discordant pairs, respectively, and let $T_x$ and $T_y$ denote the numbers of tied pairs in variables $x$ and $y$. Kendall’s $\tau$ is defined as
\begin{align}
\tau =
\frac{C - D}
{\sqrt{(C + D + T_x)(C + D + T_y)}} .
\end{align}
The coefficient satisfies $-1 \le \tau \le 1$, where $\tau = 1$ indicates identical ordering, $\tau = -1$ indicates complete inversion, and $\tau = 0$ corresponds to no monotonic association. In this work, $\tau$ is computed using scipy.stats.kendalltau, which implements the standard tie-corrected definition.

\subsubsection{Set-Based correlation measures}
\[
\begin{aligned}
E_k &:= \{\,r_1,\dots,r_k\,\}_{\text{eig}},\\
C_k^{\mathrm{CTQW}} &:= \{\,r_1,\dots,r_k\,\}_{\text{CTQW}}.
\end{aligned}
\]
We form $U = E_k \cup C_k^{\mathrm{CTQW}}$ and assign each residue $r \in U$ a rank:
\begin{align}
R_{\mathrm{eig}}(r) &=
\begin{cases}
\text{position of $r$ in }E_k, & r \in E_k, \\[2pt]
k{+}1, & r \notin E_k,
\end{cases} \nonumber \\[4pt]
R_{\mathrm{CTQW}}(r) &=
\begin{cases}
\text{position of $r$ in }C_k^{\mathrm{CTQW}}, & r \in C_k^{\mathrm{CTQW}}, \\[2pt]
k{+}1, & r \notin C_k^{\mathrm{CTQW}}.
\end{cases}
\end{align}
Items absent from a method’s top-$k$ are thus tied at rank $k{+}1$.  
We define $x_i = R_{\mathrm{eig}}(u_i)$ and $y_i = R_{\mathrm{CTQW}}(u_i)$ for $u_i \in U$.

\paragraph*{Overlap.}
Overlap@\(k\) measures the fraction of residues shared by the two top-\(k\) sets,
\begin{align}
\mathrm{Overlap@}k(E_k, C_k^{\mathrm{CTQW}})
=
\frac{|E_k \cap C_k^{\mathrm{CTQW}}|}{k}.
\end{align}
A value of \(\mathrm{Overlap@}k = 1\) indicates that the two methods identify exactly the same top-\(k\) residues, whereas \(\mathrm{Overlap@}k = 0\) indicates that the two sets share no residues.

\paragraph*{Jaccard index.}
The Jaccard index measures the similarity between the same two top-\(k\) sets by normalizing the size of their intersection by the size of their union:
\begin{align}
\mathrm{Jaccard@}k(E_k, C_k^{\mathrm{CTQW}})
=
\frac{|E_k \cap C_k^{\mathrm{CTQW}}|}
{|E_k \cup C_k^{\mathrm{CTQW}}|}.
\end{align}
A value of \(\mathrm{Jaccard@}k = 1\) indicates complete agreement between the two sets, whereas \(\mathrm{Jaccard@}k = 0\) indicates no shared residues.

\paragraph*{Interpretation.}
$\rho$ and $\tau$ capture monotonic and pairwise rank consistency (tie-aware), while Overlap and $J$ quantify direct agreement among the most central residues. Reporting all four provides a balanced assessment of both ordering and membership concordance between quantum and classical centrality measures.

\section{Results}

\subsection{Centrality correlations}

We benchmarked the CTQW centrality framework on around 150 proteins ranging from 10 to 1000 residues, covering signaling proteins, immune system proteins, myoglobin, hemoglobin, G-proteins, and tyrosine kinase receptors. This dataset size was chosen to balance structural diversity with statistical robustness, ensuring that observed correlations are not protein-specific artifacts. In representative cases, CTQW centrality highlights residues with known biological relevance, validating the method's applicability across diverse protein families. The complete list of proteins with their PDB ids and their residue counts are given in Supplementary Material 1.

\begin{table}[t]
\centering
\caption{CTQW--eigenvector correlation metrics and steady-state time $t^\star$ (Group 1).}
\label{tab:ctqw_corr_tstar_1}

\setlength{\tabcolsep}{3pt} 
\renewcommand{\arraystretch}{1.1} 

\scriptsize 

\begin{tabular}{lccccc}
\hline
PDB ID & Spearman $\rho$ & Kendall $\tau$ & Overlap & Jaccard & $t^\star$ \\
\hline
1A2B & 0.964 & 0.867 & 1.000 & 1.000 & 1187.549 \\
1AAC & 0.936 & 0.855 & 0.900 & 0.818 & 1132.422 \\
1AAY & 0.952 & 0.867 & 1.000 & 1.000 & 1352.681 \\
1ABO & 0.849 & 0.738 & 0.800 & 0.667 & 387.766 \\
1AFO & 0.976 & 0.911 & 1.000 & 1.000 & 445.769 \\
1AHD & 1.000 & 1.000 & 1.000 & 1.000 & 862.661 \\
1AKE & 0.988 & 0.956 & 1.000 & 1.000 & 1778.574 \\
1ALU & 0.964 & 0.911 & 1.000 & 1.000 & 979.541 \\
1ATP & 0.952 & 0.867 & 1.000 & 1.000 & 2761.115 \\
1AU1 & 0.976 & 0.911 & 1.000 & 1.000 & 1843.577 \\
1AVU & 0.927 & 0.822 & 1.000 & 1.000 & 1396.683 \\
1AX8 & 0.976 & 0.911 & 1.000 & 1.000 & 838.035 \\
1B0X & 0.982 & 0.927 & 0.900 & 0.818 & 464.769 \\
1B50 & 0.770 & 0.600 & 1.000 & 1.000 & 292.137 \\
1BE9 & 0.982 & 0.927 & 0.900 & 0.818 & 989.291 \\
1BFG & 0.842 & 0.733 & 1.000 & 1.000 & 454.644 \\
1BL8 & 0.945 & 0.818 & 0.900 & 0.818 & 720.905 \\
1BNR & 0.973 & 0.891 & 0.900 & 0.818 & 678.903 \\
1BTA & 1.000 & 1.000 & 1.000 & 1.000 & 567.149 \\
1BTG & 0.952 & 0.867 & 1.000 & 1.000 & 2994.250 \\
1BTL & 0.691 & 0.491 & 0.900 & 0.818 & 2716.988 \\
1BUY & 0.988 & 0.956 & 1.000 & 1.000 & 1227.426 \\
1BZ0 & 0.891 & 0.800 & 0.800 & 0.667 & 747.031 \\
1C3W & 0.976 & 0.911 & 1.000 & 1.000 & 1759.198 \\
1CD8 & 0.964 & 0.891 & 0.900 & 0.818 & 693.154 \\
1CKA & 0.927 & 0.822 & 1.000 & 1.000 & 357.890 \\
1CLB & 0.982 & 0.927 & 0.900 & 0.818 & 640.777 \\
1CRN & 0.955 & 0.855 & 0.900 & 0.818 & 398.642 \\
1CSG & 0.973 & 0.891 & 0.900 & 0.818 & 1253.552 \\
1CVJ & 0.964 & 0.891 & 0.900 & 0.818 & 1653.194 \\
1CWA & 0.903 & 0.778 & 1.000 & 1.000 & 794.283 \\
1CYC & 0.915 & 0.822 & 1.000 & 1.000 & 1218.301 \\
1D5R & 0.827 & 0.709 & 0.900 & 0.818 & 2475.728 \\
1DOK & 0.991 & 0.964 & 0.900 & 0.818 & 630.401 \\
1DOL & 0.952 & 0.867 & 1.000 & 1.000 & 1483.996 \\
1DOR & 0.964 & 0.867 & 1.000 & 1.000 & 678.278 \\
1DPX & 0.958 & 0.872 & 1.000 & 1.000 & 1459.995 \\
1DTP & 0.673 & 0.564 & 0.900 & 0.818 & 1097.420 \\
1DTX & 0.964 & 0.891 & 0.900 & 0.818 & 781.782 \\
1E44 & 0.927 & 0.822 & 0.900 & 0.818 & 525.271 \\
1EGF & 0.927 & 0.822 & 1.000 & 1.000 & 461.394 \\
1ENH & 0.988 & 0.956 & 1.000 & 1.000 & 533.522 \\
1ET1 & 1.000 & 1.000 & 1.000 & 1.000 & 832.535 \\

\hline
\end{tabular}
\end{table}

\begin{table}[t]
\centering
\caption{CTQW--eigenvector correlation metrics and steady-state time $t^\star$ (Group 2).}
\label{tab:ctqw_corr_tstar_2}

\setlength{\tabcolsep}{3pt} 
\renewcommand{\arraystretch}{1.1} 

\scriptsize 
\begin{tabular}{lccccc}
\hline
PDB ID & Spearman $\rho$ & Kendall $\tau$ & Overlap & Jaccard & $t^\star$ \\
\hline
1FD3 & 0.927 & 0.822 & 1.000 & 1.000 & 3010.000 \\
1G6G & 1.000 & 1.000 & 1.000 & 1.000 & 1140.923 \\
1G9O & 0.976 & 0.911 & 1.000 & 1.000 & 2675.362 \\
1GBQ & 0.976 & 0.911 & 1.000 & 1.000 & 1676.819 \\
1GCN & 0.916 & 0.800 & 1.000 & 1.000 & 1082.794 \\
1GFL & 0.867 & 0.778 & 1.000 & 1.000 & 1951.822 \\
1GIA & 0.976 & 0.911 & 1.000 & 1.000 & 1266.053 \\
1GPC & 0.976 & 0.911 & 1.000 & 1.000 & 1126.922 \\
1GRM & 0.976 & 0.929 & 1.000 & 1.000 & 2409.395 \\
1GZN & 0.952 & 0.867 & 1.000 & 1.000 & 2822.621 \\
1H68 & 0.915 & 0.800 & 1.000 & 1.000 & 1949.697 \\
1HGU & 0.964 & 0.867 & 1.000 & 1.000 & 1411.684 \\
1HRC & 0.811 & 0.693 & 0.700 & 0.538 & 423.767 \\
1HRH & 0.782 & 0.673 & 0.900 & 0.818 & 972.416 \\
1HRJ & 0.988 & 0.956 & 1.000 & 1.000 & 1493.247 \\
1HUM & 0.964 & 0.891 & 0.900 & 0.818 & 1118.796 \\
1I1B & 0.952 & 0.867 & 1.000 & 1.000 & 772.532 \\
1IDB & 0.927 & 0.822 & 1.000 & 1.000 & 648.777 \\
1IEP & 0.964 & 0.911 & 1.000 & 1.000 & 640.902 \\
1IGD & 0.964 & 0.867 & 1.000 & 1.000 & 588.525 \\
1IHB & 0.854 & 0.744 & 0.900 & 0.818 & 898.040 \\
1IO2 & 0.964 & 0.891 & 0.900 & 0.818 & 606.774 \\
1IRO & 0.806 & 0.689 & 1.000 & 1.000 & 2370.017 \\
1ITF & 0.964 & 0.867 & 1.000 & 1.000 & 694.779 \\
1J00 & 0.661 & 0.539 & 0.900 & 0.818 & 795.033 \\
1J7Z & 0.952 & 0.867 & 1.000 & 1.000 & 2811.745 \\
1J8R & 0.982 & 0.927 & 0.900 & 0.818 & 679.653 \\
1J9O & 1.000 & 1.000 & 1.000 & 1.000 & 595.900 \\
1JHH & 0.855 & 0.738 & 1.000 & 1.000 & 522.646 \\
1JM7 & 0.927 & 0.822 & 1.000 & 1.000 & 977.416 \\
1JMQ & 0.927 & 0.822 & 1.000 & 1.000 & 835.160 \\
1KLC & 0.939 & 0.863 & 0.800 & 0.667 & 630.401 \\
1KZK & 1.000 & 1.000 & 1.000 & 1.000 & 875.786 \\
1L2Y & 1.000 & 1.000 & 1.000 & 1.000 & 353.640 \\
1LMB & 0.964 & 0.891 & 0.900 & 0.818 & 766.906 \\
1M14 & 0.582 & 0.492 & 0.800 & 0.667 & 808.159 \\
1M8A & 0.988 & 0.956 & 1.000 & 1.000 & 1197.425 \\
1MAI & 0.988 & 0.956 & 1.000 & 1.000 & 565.024 \\
1MBN & 0.891 & 0.778 & 1.000 & 1.000 & 890.540 \\
1MJC & 0.982 & 0.927 & 0.900 & 0.818 & 715.030 \\
1MLR & 0.964 & 0.891 & 0.900 & 0.818 & 920.541 \\
1NCO & 0.976 & 0.911 & 1.000 & 1.000 & 2927.122 \\
1O7Y & 0.952 & 0.867 & 1.000 & 1.000 & 2917.247 \\
\hline
\end{tabular}
\end{table}

\begin{table}[t]
\centering
\caption{CTQW--eigenvector correlation metrics and steady-state time $t^\star$ (Group 3).}
\label{tab:ctqw_corr_tstar_3}

\setlength{\tabcolsep}{3pt} 
\renewcommand{\arraystretch}{1.1} 

\scriptsize 
\begin{tabular}{lccccc}
\hline
PDB ID & Spearman $\rho$ & Kendall $\tau$ & Overlap & Jaccard & $t^\star$ \\
\hline
1OT8 & 0.952 & 0.867 & 1.000 & 1.000 & 1045.293 \\
1P7D & 0.976 & 0.911 & 1.000 & 1.000 & 2385.394 \\
1PAZ & 0.976 & 0.911 & 1.000 & 1.000 & 1041.918 \\
1PG1 & 0.982 & 0.927 & 0.900 & 0.818 & 385.891 \\
1PGA & 0.988 & 0.956 & 1.000 & 1.000 & 345.765 \\
1PLC & 0.988 & 0.956 & 1.000 & 1.000 & 2970.624 \\
1PNJ & 0.927 & 0.822 & 1.000 & 1.000 & 414.267 \\
1PTQ & 0.915 & 0.822 & 1.000 & 1.000 & 322.139 \\
1Q5O & 0.952 & 0.867 & 1.000 & 1.000 & 2587.900 \\
1QAU & 0.982 & 0.927 & 0.900 & 0.818 & 1162.673 \\
1QJP & 0.962 & 0.891 & 0.900 & 0.818 & 724.780 \\
1QUQ & 0.952 & 0.867 & 1.000 & 1.000 & 1764.823 \\
1QZ0 & 0.988 & 0.956 & 1.000 & 1.000 & 1284.054 \\
1RCY & 0.891 & 0.738 & 0.800 & 0.667 & 930.416 \\
1RNH & 0.952 & 0.867 & 1.000 & 1.000 & 798.283 \\
1ROP & 0.866 & 0.758 & 1.000 & 1.000 & 279.886 \\
1ROO & 0.982 & 0.927 & 0.900 & 0.818 & 1276.553 \\
1RW5 & 0.988 & 0.956 & 1.000 & 1.000 & 1214.926 \\
1RX2 & 0.982 & 0.927 & 0.900 & 0.818 & 2144.768 \\
1RZX & 0.976 & 0.911 & 1.000 & 1.000 & 1031.418 \\
1SDF & 0.964 & 0.891 & 0.900 & 0.818 & 1005.541 \\
1SHF & 0.918 & 0.745 & 0.900 & 0.818 & 2866.373 \\
1SJQ & 0.927 & 0.822 & 1.000 & 1.000 & 604.649 \\
1SPS & 0.976 & 0.911 & 1.000 & 1.000 & 1000.417 \\
1SRL & 0.927 & 0.822 & 1.000 & 1.000 & 952.792 \\
1STN & 0.940 & 0.863 & 0.800 & 0.667 & 601.274 \\
1TBP & 0.939 & 0.863 & 0.800 & 0.667 & 1255.677 \\
1TEN & 0.915 & 0.822 & 1.000 & 1.000 & 378.641 \\
1THX & 0.952 & 0.867 & 1.000 & 1.000 & 2038.765 \\
1TNF & 0.927 & 0.822 & 1.000 & 1.000 & 1149.548 \\
1TRZ & 0.915 & 0.822 & 1.000 & 1.000 & 457.769 \\
1TU4 & 0.982 & 0.927 & 0.900 & 0.818 & 746.281 \\
1TZE & 0.976 & 0.911 & 1.000 & 1.000 & 688.154 \\
1UBQ & 0.952 & 0.867 & 1.000 & 1.000 & 497.396 \\
1UNQ & 0.964 & 0.891 & 0.900 & 0.818 & 1386.183 \\
1URN & 0.964 & 0.891 & 0.900 & 0.818 & 1100.045 \\
1VII & 0.945 & 0.855 & 0.900 & 0.818 & 431.143 \\
1VPU & 0.982 & 0.927 & 0.900 & 0.818 & 952.042 \\
1W4E & 0.915 & 0.822 & 1.000 & 1.000 & 1327.429 \\
1WZ2 & 0.867 & 0.778 & 1.000 & 1.000 & 362.390 \\
1XNB & 0.945 & 0.818 & 0.900 & 0.818 & 2298.470 \\
1XY1 & 0.976 & 0.929 & 1.000 & 1.000 & 448.394 \\
1YGM & 0.988 & 0.956 & 1.000 & 1.000 & 266.511 \\
\hline
\end{tabular}
\end{table}

\begin{table}[t]
\centering
\caption{CTQW--eigenvector correlation metrics and steady-state time $t^\star$ (Group 4).}
\label{tab:ctqw_corr_tstar_4}

\setlength{\tabcolsep}{3pt} 
\renewcommand{\arraystretch}{1.1} 

\scriptsize 
\begin{tabular}{lccccc}
\hline
PDB ID & Spearman $\rho$ & Kendall $\tau$ & Overlap & Jaccard & $t^\star$ \\
\hline
1YWO & 0.982 & 0.927 & 0.900 & 0.818 & 2211.143 \\
256B & 0.955 & 0.855 & 0.900 & 0.818 & 1487.372 \\
2AAS & 0.988 & 0.956 & 1.000 & 1.000 & 613.524 \\
2ABD & 0.945 & 0.818 & 0.900 & 0.818 & 1419.184 \\
2CBA & 0.903 & 0.778 & 1.000 & 1.000 & 1304.555 \\
2CHF & 0.915 & 0.822 & 1.000 & 1.000 & 806.034 \\
2CHT & 0.982 & 0.927 & 0.900 & 0.818 & 1113.171 \\
2CI2 & 0.976 & 0.911 & 1.000 & 1.000 & 543.273 \\
2CQV & 0.903 & 0.778 & 1.000 & 1.000 & 466.144 \\
2CRD & 0.982 & 0.927 & 0.900 & 0.818 & 568.274 \\
2CRO & 0.964 & 0.891 & 0.900 & 0.818 & 474.519 \\
2GLS & 0.952 & 0.867 & 1.000 & 1.000 & 916.666 \\
2GLH & 0.903 & 0.778 & 1.000 & 1.000 & 747.031 \\
2HHB & 0.945 & 0.818 & 0.900 & 0.818 & 1821.076 \\
2IG2 & 0.964 & 0.891 & 0.900 & 0.818 & 1678.944 \\
2K6O & 0.927 & 0.822 & 1.000 & 1.000 & 730.406 \\
2LHB & 0.976 & 0.911 & 1.000 & 1.000 & 649.902 \\
2MAG & 0.952 & 0.867 & 1.000 & 1.000 & 816.159 \\
2MLT & 0.830 & 0.644 & 1.000 & 1.000 & 375.266 \\
2OZF & 0.988 & 0.956 & 1.000 & 1.000 & 1316.431 \\
2PDZ & 0.952 & 0.867 & 1.000 & 1.000 & 1101.170 \\
2RH1 & 0.976 & 0.911 & 1.000 & 1.000 & 2521.774 \\
2RLF & 0.964 & 0.891 & 0.900 & 0.818 & 928.666 \\
3B5D & 0.952 & 0.867 & 1.000 & 1.000 & 995.666 \\
3EML & 0.964 & 0.891 & 0.900 & 0.818 & 1442.810 \\
3GB1 & 0.964 & 0.891 & 0.900 & 0.818 & 574.649 \\
3IL8 & 0.982 & 0.927 & 0.900 & 0.818 & 1498.872 \\
3UG9 & 0.952 & 0.867 & 1.000 & 1.000 & 1890.200 \\
3UON & 0.964 & 0.891 & 0.900 & 0.818 & 1203.800 \\
4AZU & 0.991 & 0.964 & 0.900 & 0.818 & 278.761 \\
4FXC & 0.879 & 0.767 & 0.900 & 0.818 & 874.661 \\
4GCR & 0.964 & 0.891 & 0.900 & 0.818 & 540.273 \\
4HHB & 0.927 & 0.822 & 1.000 & 1.000 & 1112.421 \\
4IB4 & 0.964 & 0.891 & 0.900 & 0.818 & 823.410 \\
4INS & 0.842 & 0.644 & 1.000 & 1.000 & 354.765 \\
4X1H & 0.973 & 0.891 & 0.900 & 0.818 & 2275.970 \\
5DU1 & 0.952 & 0.867 & 1.000 & 1.000 & 483.145 \\
5P21 & 0.988 & 0.956 & 1.000 & 1.000 & 1329.555 \\
5PTI & 0.952 & 0.867 & 1.000 & 1.000 & 754.031 \\
5Y9Z & 0.964 & 0.867 & 1.000 & 1.000 & 1636.318 \\
6LYZ & 0.881 & 0.738 & 0.800 & 0.667 & 779.532 \\
6RLX & 0.964 & 0.867 & 1.000 & 1.000 & 293.887 \\
7RSA & 0.991 & 0.964 & 0.900 & 0.818 & 723.905 \\
\hline
\end{tabular}
\end{table}

Tables I–IV list rank and set‐based correlations between CTQW and eigenvector centralities for all proteins, arranged alphabetically for readability.
The correlations are consistently high, confirming strong monotonic agreement and overlap among top-ranked residues.

 \begin{figure}[t]
  \centering
  \includegraphics[width=0.7\linewidth]{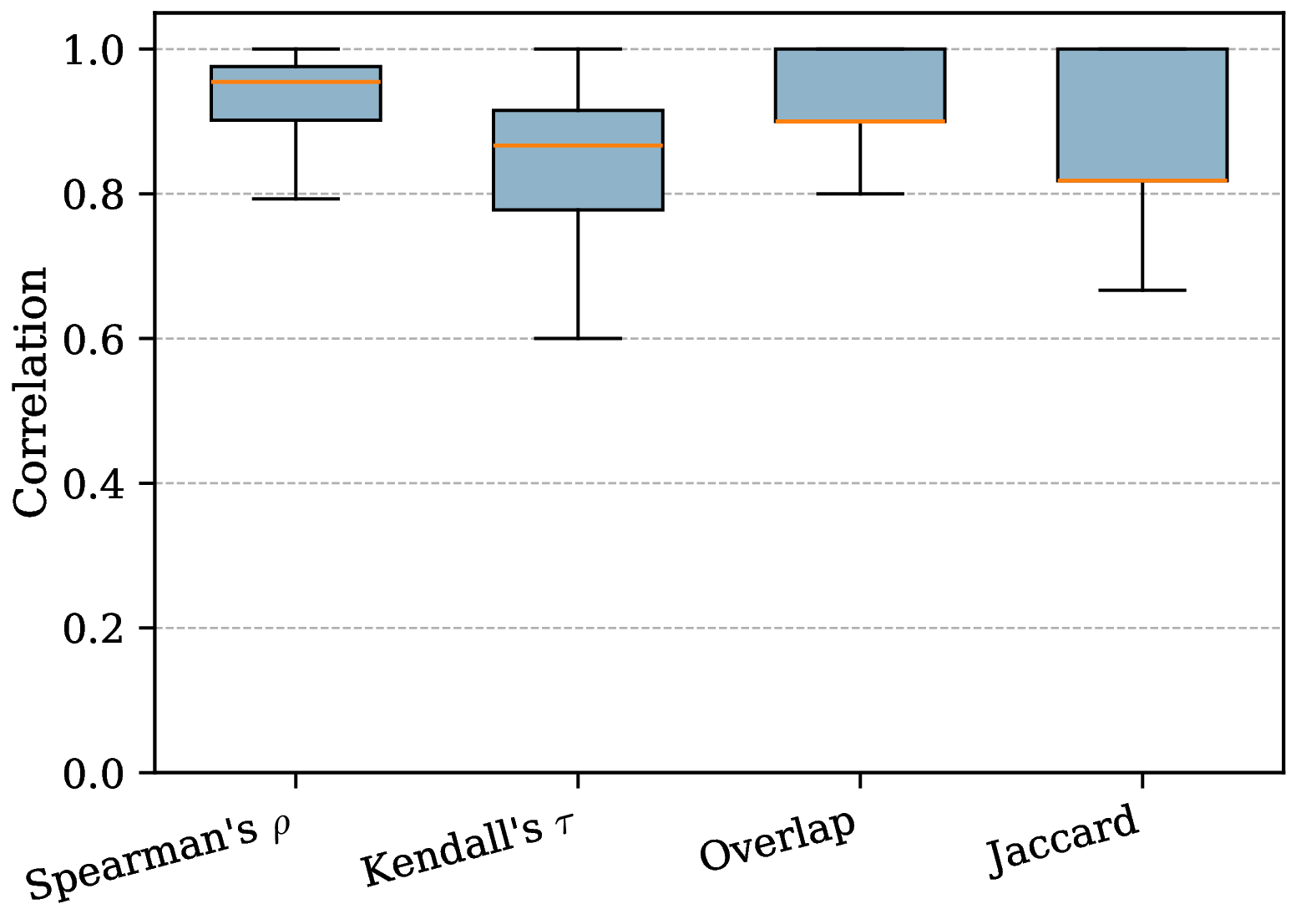}
  \caption{Box plot comparison of the four correlation and overlap metrics: Spearman’s $\rho$, Kendall’s $\tau$, Overlap, and Jaccard across all analyzed proteins. Each distribution summarizes the consistency between CTQW and eigenvector centralities. The high median values and narrow interquartile ranges indicate strong overall agreement.}
  \label{fig:boxplot}
\end{figure}

\subsection{Spectral Gap: Quantum vs.\ Classical Transition Operators}

\begin{figure}[t]
    \centering
    \includegraphics[width=0.7\linewidth]{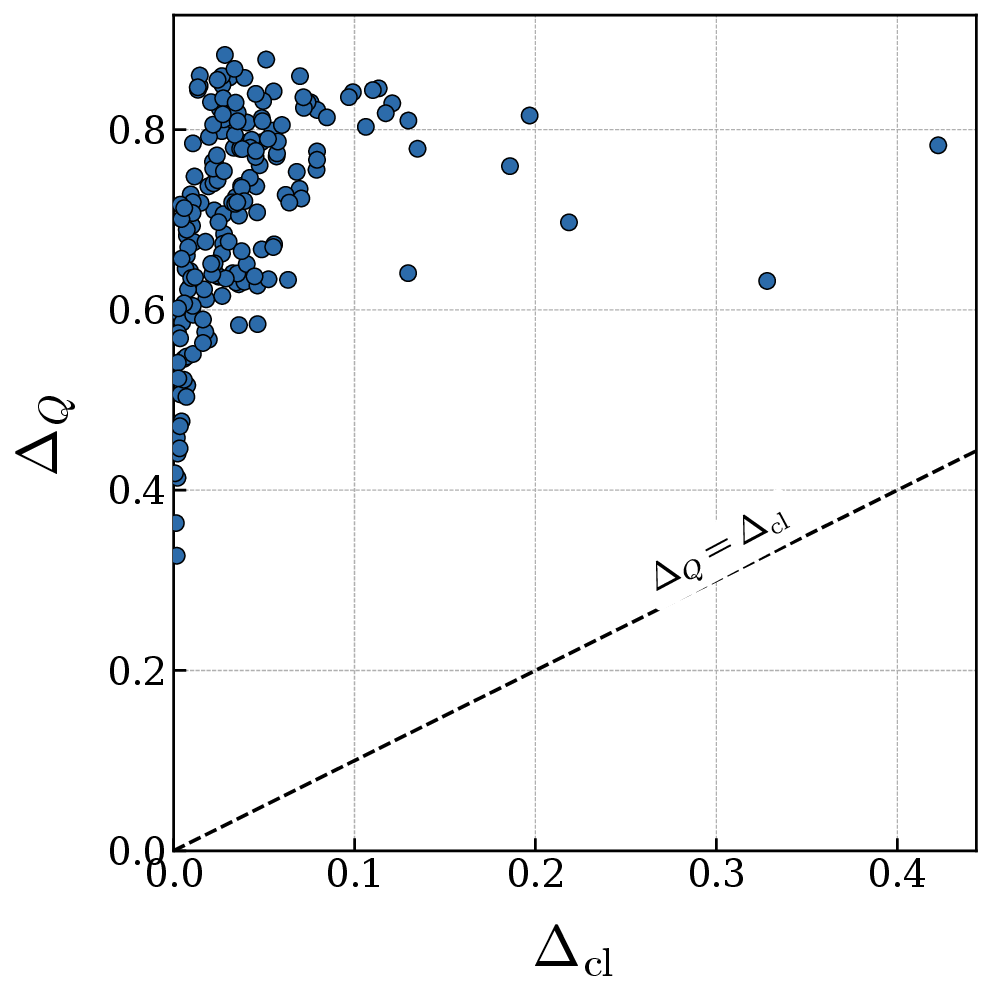}
    \caption{Comparison of quantum and classical spectral gaps across the protein dataset. Each point corresponds to a residue interaction network. The dashed line denotes $\Delta_Q = \Delta_{\mathrm{cl}}$. Nearly all points lie above the diagonal, indicating that the dephased quantum transition map is spectrally more concentrated than the classical diffusion operator.}
    \label{fig:spectral_gap_scatter}
\end{figure}

Figure~\ref{fig:spectral_gap_scatter} compares $\Delta_Q$ and $\Delta_{\mathrm{cl}}$ across the protein dataset. Classical gaps cluster in the regime $\Delta_{\mathrm{cl}} \ll 1$, reflecting slow convergence of the diffusive random walk due to sparse connectivity and structural bottlenecks. Quantum gaps are consistently larger, with most values in the range $0.6$--$0.85$, and all proteins lie well above the diagonal $\Delta_Q = \Delta_{\mathrm{cl}}$.

\begin{figure}[t]
    \centering
    \includegraphics[width=0.7\linewidth]{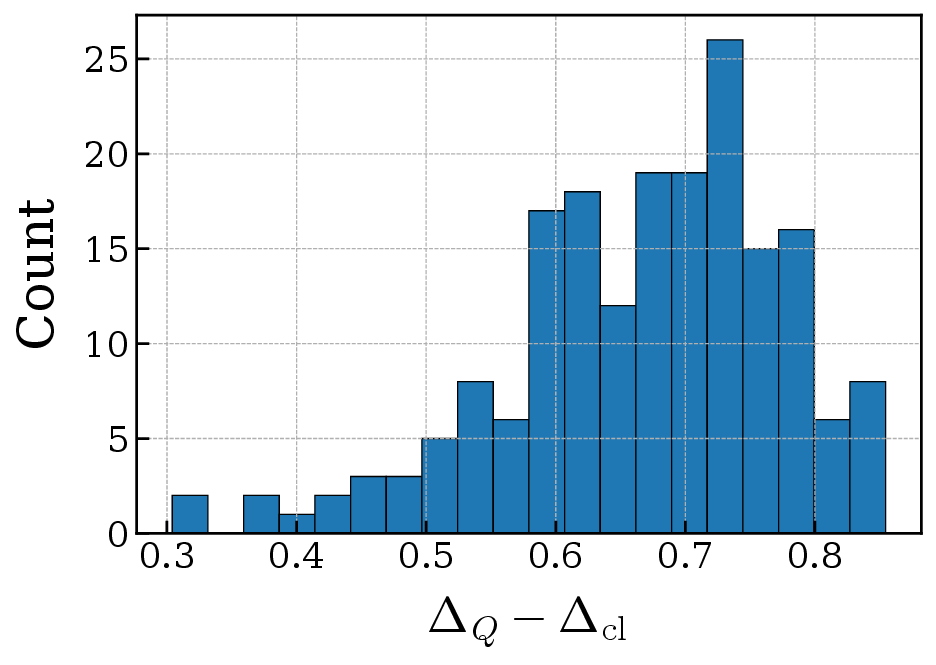}
    \caption{Distribution of the spectral gap difference $\Delta_Q - \Delta_{\mathrm{cl}}$. The distribution is sharply peaked in the range $0.6$--$0.75$, with all values positive, indicating that the quantum walk consistently exhibits a larger spectral gap than the classical random walk across all proteins studied. }
    \label{fig:spectral_gap_hist}
\end{figure}

The distribution of the difference $\Delta_Q - \Delta_{\mathrm{cl}}$, shown in Fig.~\ref{fig:spectral_gap_hist}, is sharply peaked around $0.6$--$0.75$ with all values positive, confirming that this separation is systematic across diverse protein topologies rather than confined to a particular structural class.

A larger spectral gap for the quantum walk relative to the classical random walk implies an increased separation between the eigenvalue associated with the steady-state subspace and the next dominant eigenvalue. This enhanced separation suppresses subdominant spectral contributions more efficiently. As the mixing time is provably inversely proportional to the spectral gap, this directly implies faster convergence of the quantum walk dynamics.\cite{PhysRevA.102.022423}

\subsection{Steady-State Convergence Times Across Proteins}

Beyond the finite-time averaging used for centrality extraction, we previously derived an exact analytic characterization of the infinite-time steady state of the CTQW dynamics on each residue interaction network in Sec ~\nameref{sec:ctqw_steady_state}. With the Hamiltonian identified as the symmetric adjacency matrix $H = A$, the steady-state distribution follows directly from the spectral decomposition.
\begin{equation}
    A = V \Lambda V^{\top}.
\end{equation}

In the long-time limit, phase coherence between distinct eigenvalues averages out, while contributions within degenerate eigenspaces persist. This spectral dephasing yields a closed-form expression for the steady-state probability vector $\bar{P}$, independent of time discretization, providing a rigorous benchmark for numerical and hardware implementations.

To quantify the dynamical stabilization timescale, we computed the running long time mean
\begin{equation}
\bar{P}(T) = \frac{1}{T} \int_0^T P(t)\, dt,
\end{equation}
from explicit unitary evolution $e^{-iAt}$ and monitored its convergence toward $\bar{P}(\infty)$ using the $L^1$ distance
\begin{equation}
\|\bar{P}(T) - \bar{P}(\infty)\|_1.
\end{equation}
The convergence time $T^*$ was defined as the smallest $T$ for which this distance falls below a chosen tolerance. This establishes a direct connection between the spectral structure of the adjacency matrix and the dynamical timescale of coherent quantum propagation on protein networks.

To quantify the time required for the quantum walks to reach steady state on residue interaction networks, we computed the convergence time $t^*$ for each protein, defined as the first time at which the running long time average Ces\`aro mean falls below the prescribed tolerance relative to its analytical steady-state value. This quantity provides a physically interpretable timescale for the decay of interference-driven oscillations and the emergence of a stable long-time occupation profile.

\begin{figure}[t]
  \centering
  \includegraphics[width=0.7\linewidth]{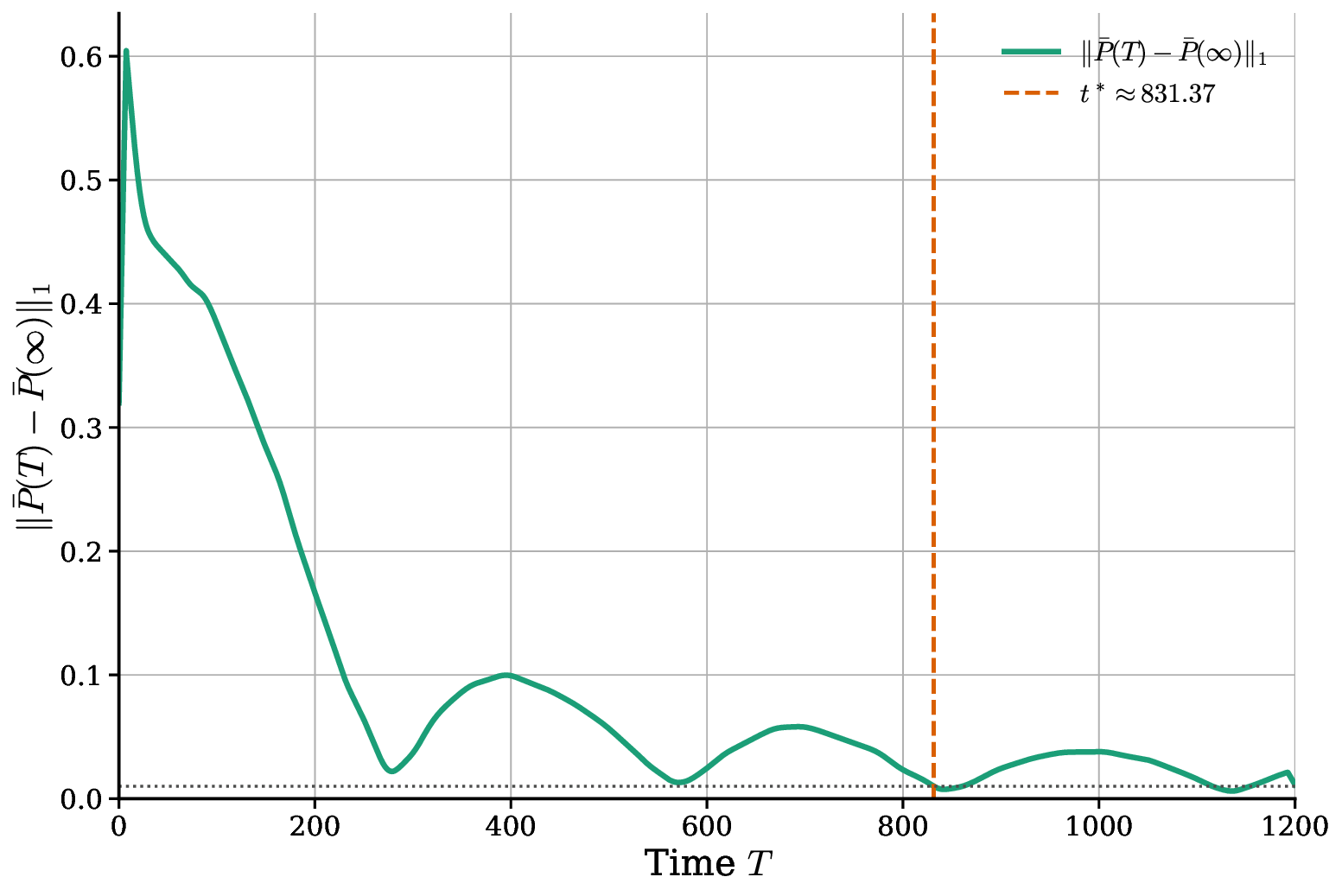}
  \caption{Validation of steady-state convergence for the CTQW centrality measure on the 1IJZ residue interaction network. The figure shows the $L_1$-error $\|\bar{P}(T)-\bar{P}(\infty)\|_1$ between the finite-time C\'esaro mean and the analytic infinite-time distribution. The detected convergence time $t^*$ (vertical dashed line) marks the earliest time horizon at which the error remains below tolerance $\varepsilon$, confirming that finite-time averaging accurately approximates the steady state.}
  \label{fig:steady}
\end{figure}

To verify that the finite-time C\'esaro averaging procedure faithfully approximates the analytic steady state, we monitor the $L_1$-distance $\|\bar{P}(T)-\bar{P}(\infty)\|_1$ as a function of the time horizon $T$. The resulting convergence profile for a representative protein (1IJZ) is presented in Fig.~\ref{fig:steady}. 

\subsection{Proof-of-principle hardware demonstration}
To assess experimental feasibility, the CTQW–centrality circuit was executed on the \texttt{ibm\_kingston}\cite{javadiabhari2024quantumcomputingqiskit} superconducting processor via the Qiskit Runtime interface integrated with PennyLane. 
Each circuit was sampled with 1024 shots, and long-time-averaged probabilities
\(\bar P_i\) yielded the CTQW centralities.

\begin{table}[t]
\centering
\caption{\textbf{Residue ranking for oxytocin (PDB 1XY1).}
Ranks are listed top-to-bottom (highest $\rightarrow$ lowest).}
\vspace{4pt}
\footnotesize
\setlength{\tabcolsep}{4pt}
\renewcommand{\arraystretch}{1.05}
\begin{tabular}{cccc}
\hline\hline
\textbf{Rank} & \textbf{Classical} & \textbf{Simulator} & \textbf{IBM Kingston} \\
\hline
1 & 6 & 6 & 6 \\
2 & 5 & 5 & 5 \\
3 & 1 & 2 & 1 \\
4 & 2 & 1 & 2 \\
5 & 3 & 3 & 4 \\
6 & 7 & 4 & 3 \\
7 & 4 & 7 & 7 \\
8 & 8 & 8 & 8 \\
9 & 9 & 9 & 9 \\
\hline\hline
\end{tabular}
\label{tab:device-comparison}
\end{table}

As a representative case, the nine-residue oxytocin network (PDB 1XY1) was analyzed across classical, simulator, and hardware implementations. Despite noise in the device runs, the highest-ranked residues remain consistent, demonstrating that CTQW-derived centralities are experimentally accessible on current noisy intermediate-scale quantum hardware. The hardware results reproduce the simulator’s ranking profile within statistical uncertainty, with only minor permutations among the upper-rank residues. Such deviations stem from finite-shot sampling and two-qubit gate infidelities in the compiled dense-unitary representation of \(U(t)\).  Future work should employ structure-preserving Hamiltonian simulation (e.g., Trotter–Suzuki \cite{trot24, aplQ, Zhao2023, Lee_2023} or sparse-encoding schemes\cite{Yang2025dictionarybased}) to reduce circuit depth while maintaining the same CTQW averaging protocol.

\section{Biological Validation via Functional Residue Identification}
Within the CTQW framework, probability amplitude propagation can be interpreted as coherent transport of structural influence across the protein contact network. Interference between multiple interaction pathways modulates how strongly perturbations originating at one residue affect distant regions, providing a dynamical analogue of allosteric communication.

In a broader biological context, residues occupying central positions in residue interaction networks often correspond to key sites governing protein stability, folding, and function. High-centrality residues serve as topological and energetic hubs through which structural perturbations or allosteric signals propagate.~\cite{Atilgan2004,Yang2009-mh,10.1093/nar/gkn433} Such residues frequently coincide with catalytic centers, metal-binding motifs, or interface residues at ligand- or receptor-binding sites, where even minor perturbations can lead to large-scale functional changes~\cite{Chea2007-ud,Negre2018,Reetz2009-dm, Süel2003}.
Centrality analysis therefore provides a quantitative route to link network topology with biochemical activity, allowing identification of residues whose spatial position and connectivity confer disproportionate influence on the overall dynamics of the protein~\cite{AMITAI20041135, Yang2009-mh}. The strong agreement between CTQW-derived rankings and experimental evidence across representative systems demonstrates that quantum-walk–based centrality is not merely a mathematical abstraction, but a physically grounded predictor of functional importance, offering a promising computational avenue for mapping active and allosteric sites directly from structural data.

To assess whether CTQW centrality identifies biologically meaningful residues, we benchmarked it against two representative systems: protein kinase A, for which experimentally established key residues are available from prior centrality studies, and oxytocin, a compact peptide with well-characterized functional residues.

\subsection{Recovery of Experimentally Known Functional Residues in Protein Kinase A}

To evaluate whether the proposed centrality measures capture biologically
important sites, we follow the benchmarking protocol of Kornev~\textit{et al.}\cite{doi:10.1073/pnas.2215420119}, who defined a reference set of 26 experimentally established functional residues in protein kinase A (PKA) and used it to assess the discriminative power of different protein residue network (PRN) metrics. The authors introduced a key score defined as the fraction of total centrality concentrated on these benchmark residues. If $c_i$ denotes the centrality of residue $i$ and $K$ denotes the set of 26 key residues, the key score is defined as

\begin{equation}
\mathrm{KeyScore}
=
\frac{\sum_{i \in K} c_i}{\sum_{j \in V} c_j},
\end{equation}

where $V$ denotes the set of all residues in the protein.

For PKA, which contains 336 residues, random selection of 26 residues would yield a key score of approximately $\frac{26}{336} \approx 0.08$ ($\sim 8\%$), establishing the random baseline against which centrality measures are evaluated. Therefore, centrality metrics that identify biologically important residues should produce key scores significantly larger than $8\%$.

In the present work we computed the key score using the same definition. Both eigenvector and CTQW centrality vectors were normalized such that $\sum_i c_i = 1$, so that the key score reduces to $100\sum_{i \in K} c_i$.

Using this benchmark, eigenvector centrality produced a key score of $15.21\%$, while CTQW centrality yielded
$16.75\%$. Both values substantially exceed the random baseline, demonstrating that the centrality measures preferentially concentrate importance on biologically relevant residues.

These values can be compared directly with the results reported by Kornev et al., summarized in Fig.~2 of their work. That study examined four classical centrality measures: degree centrality (DC), betweenness centrality (BC), closeness centrality (CL), and eigenvector centrality (EG), applied to several types of PRNs constructed from molecular dynamics trajectories. Their results showed that closeness centrality performed essentially at the random baseline ($\sim 8$--$9\%$), indicating poor ability to distinguish functionally important residues. Degree centrality and eigenvector centrality produced moderate enrichment with
key scores typically around $14$--$17\%$ depending on the network construction method.

The highest performance reported in that study was obtained
for betweenness centrality applied to LSP-based residue interaction networks, which achieved key scores approaching $25$--$30\%$.
With the exception of this LSP–BC case, the key scores obtained in the present work for eigenvector centrality ($15.21\%$) and CTQW centrality ($16.75\%$) are comparable to or higher than those obtained for the other classical centrality measures considered in the original study.

In addition to the key-score comparison, we examined the rank-based recovery of the benchmark residues. Among the top 26 residues ranked by centrality, eigenvector centrality identifies 7 of the 26 benchmark residues, while CTQW centrality identifies 8 of the 26, corresponding to detection rates of approximately $27\%$ and $31\%$, respectively. These values again substantially exceed the random expectation of $26/336 \approx 8\%$.

Furthermore, the rankings produced by eigenvector and CTQW centrality are strongly correlated. Among the highest-ranked residues, 18 of the top 20 and 23 of the top 26 residues are shared between the two rankings.
This high overlap indicates that the long-time CTQW steady state largely reflects the spectral structure of the residue interaction network while incorporating contributions from multiple eigenmodes through coherent quantum transport.

Taken together, these results demonstrate that both spectral and quantum-walk-based centrality measures are capable of recovering biologically important residues associated with the catalytic core of protein kinase A. Notably, the CTQW centrality achieves enrichment comparable to the best-performing classical spectral measures and exceeds most classical centrality metrics considered in the original study, with the only consistently stronger method being betweenness centrality on LSP-based networks.

\subsection{Experimental Validation for Oxytocin (1XY1).}

To evaluate the biological validity of the quantum-walk framework, we analyzed the nonapeptide hormone oxytocin (PDB ID: 1XY1), whose compact and well-characterized structure provides a stringent benchmark for residue centrality analysis. The CTQW-based centrality measure identified Tyr$^{2}$, Ile$^{3}$, Asn$^{5}$, and Cys$^{6}$ as the most central residues within the residue–interaction network. Notably, these residues coincide with functionally crucial sites established through experimental studies. Spectroscopic and receptor-binding analyses have shown that Tyr$^{2}$ and Cys$^{6}$ are essential for receptor activation and ligand recognition~\cite{Liu, Gimpl2001-sd}, while molecular dynamics and ion-mobility mass spectrometry measurements demonstrated that Tyr$^{2}$, Ile$^{3}$, and Cys$^{6}$ engage in Zn$^{2+}$ coordination and exhibit restricted backbone flexibility, indicative of their structural importance~\cite{md_ot_zinc, ims_ot_bd, Dyson2005-aa}.
The strong correspondence between residues predicted by CTQW centrality and experimentally validated functional determinants highlights the capacity of quantum-walk dynamics to identify biologically meaningful nodes in protein networks. Unlike conventional eigenvector-based approaches that depend solely on the static topology of the graph, CTQW centrality encodes the effects of coherent propagation and interference, allowing it to capture collective structural roles of residues that mediate stability and communication across the molecular scaffold.

\section{Discussion}

Across the full protein dataset, CTQW centrality exhibits consistently strong agreement with eigenvector centrality, indicating that long-time averaged quantum-walk dynamics effectively capture the dominant spectral structure of residue interaction networks. In the majority of proteins, both rank-based metrics (Spearman's $\rho$, Kendall's $\tau$) and set-based metrics (Overlap and Jaccard) attain high values, demonstrating that CTQW preserves both the global ordering and the identity of the most central residues. This agreement is particularly pronounced in structurally compact and well-connected proteins, where the network is governed by a small number of leading eigenmodes. In such cases, the CTQW steady state closely mirrors the Perron eigenvector, resulting in near-perfect correspondence between quantum and classical centrality rankings. These systems represent a regime in which our quantum walk treatment and classical spectral structure effectively align. Systematic deviations from perfect agreement arise in proteins with irregular topology, reduced connectivity, or heterogeneous structural organization. In these networks, multiple eigenmodes contribute comparably to the dynamics, and quantum interference redistributes probability among closely ranked residues. Consequently, small but noticeable differences emerge in pairwise rankings---most clearly reflected in Kendall's $\tau$, while the highest-centrality residues remain largely unchanged. This behavior indicates that CTQW centrality retains robustness at the top of the ranking while introducing enhanced sensitivity to finer structural variations.

The box-plot analysis in Fig.~\ref{fig:boxplot} provides a global statistical summary of these trends across all proteins. All four metrics are strongly concentrated near their upper bounds, confirming that the agreement between CTQW and eigenvector centralities is both high and consistent across diverse network topologies. Spearman's $\rho$ exhibits the tightest clustering near unity, indicating that overall rank ordering is preserved with high fidelity, whereas Kendall's $\tau$ shows a slightly broader distribution, reflecting localized pairwise discrepancies induced by interference among nearly degenerate eigenstates. The set-based metrics further reinforce this picture. Overlap remains consistently high, indicating that most top-ranked residues are shared between the two centrality measures. The Jaccard index displays a broader spread due to its stricter normalization, which penalizes mismatches more strongly when residues enter or leave the top-$k$ sets. Nevertheless, its overall distribution confirms substantial agreement in identifying the most important residues. Crucially, the interquartile ranges remain narrow across all metrics, indicating that the agreement between CTQW and eigenvector centrality is robust to variation in protein size, fold, and connectivity. The lower tails observed in Kendall's $\tau$ and Jaccard can be attributed to proteins with elongated or sparsely connected structures, where coherent dynamics amplify subtle structural asymmetries without significantly altering the dominant centrality landscape. These results support the interpretation of CTQW centrality as a dynamical generalization of eigenvector centrality. 

\section{Conclusion}

While eigenvector centrality is determined by the leading spectral mode, CTQW centrality emerges from coherent evolution under the network Hamiltonian and therefore incorporates contributions from multiple eigenmodes. This produces a steady-state distribution that preserves the principal spectral structure while encoding additional dynamical information about the network. Given their strong agreement, a natural question is whether CTQW centrality potentially offers any practical advantages over traditional eigenvector centrality calculations. In that regard, we must emphasize that the value of the framework lies not in replacing classical measures, but in providing a physically motivated generalization rooted in quantum-walks that has the potential to account for features that traditional centrality measures fail to capture. 

The CTQW formulation naturally weights contributions from multiple eigenmodes through interference, offering a richer characterization of network structure that becomes particularly relevant in proteins with near-degenerate spectra or heterogeneous connectivity. Crucially, the formulation maps directly onto quantum hardware via Hamiltonian simulation, and logarithmic qubit scaling, requiring only $q = \lceil \log_2 n \rceil$ qubits for an $n$-residue protein, making even large systems accessible on modest quantum registers, since a 512-residue protein would require only nine qubits. Combined with sparse Hamiltonian simulation algorithms whose complexity is polylogarithmic in system size, this encoding offers an asymptotically favorable route to analyzing residue interaction networks that may become impractical to diagonalize classically as system sizes grow into the thousands of residues or when ensemble-level analyses demand repeated centrality evaluations across large numbers of conformational snapshots.

The larger spectral gaps observed for the quantum walk indicate a greater separation between the steady-state eigenvalue and the next dominant mode. Since the mixing time scales inversely with the spectral gap, this implies faster convergence compared to classical diffusion. In contrast, smaller classical gaps reflect diffusive bottlenecks that slow probability redistribution. Overall, the quantum framework exhibits faster stabilization and provides a more efficient dynamical view of residue interaction topology. We note that the present analysis uses a single contact definition (C$_\alpha$ distance $< 8$\,\AA, inverse-square weighting); future work should assess sensitivity to alternative cutoffs and weighting schemes to establish the robustness of the centrality measures further. Other natural extensions include applying the framework to allosteric pathway prediction, mutational sensitivity mapping, and ensemble-averaged networks from molecular dynamics trajectories.

We establish continuous-time quantum-walk centrality as a principled and computationally tractable framework for analyzing protein residue interaction networks. By bridging quantum-walk dynamics with spectral graph theory, the approach introduces sensitivity to finer variations in structure through multi-eigenmode interference, while reproducing established structural signatures. As quantum hardware continues to scale, logarithmic qubit encoding positions CTQW centrality as a prime candidate towards realising quantum-native analysis of large biomolecular networks \cite{yazdi25, shehab24}.

\section*{Code Availability}
The code supporting this study is available at \url{https://github.com/IshmamShah/CTQW_RIN}.

\section*{Acknowledgments}

This work was supported by the U.S. Department of Energy (DOE) Office of Basic Energy Sciences, Award No. DE-SC0026309. SIM acknowledges discussions with Dr. Farhan T. Chowdhury and Dr. Stephen Wolfram. 

\section*{Supporting information}


The following files are available free of charge.
\begin{itemize}
  \item Supplementary Information 1: The comprehensive list of proteins, their corresponding PDB identifiers and their residue numbers.

\end{itemize}


\bibliography{references.bib}




  

\end{document}